\documentclass[prl,twocolumn,showpacs,amsmath,amssymb,superscriptaddress]{revtex4}

\usepackage[dvips]{graphicx}

\usepackage{graphics}

\usepackage{dcolumn}

\usepackage{bm}

\usepackage[usenames]{color}

\begin{document}

\title{Domain walls in (Ga,Mn)As diluted magnetic semiconductor}

\author{Akira~Sugawara}

\affiliation{Initial Research Project, Okinawa Institute of Science and Technology, c/o Hitachi Advanced Research Laboratory, Akanuma 2520, Hatoyama, Saitama 350-0395, Japan}
 \altaffiliation[Present address: ]{Advanced Research Laboratory, Hitachi Ltd.}
\email{akira.sugawara.ne@hitachi.com}
\author{H.~Kasai}

\affiliation{Advanced Research Laboratory, Hitachi Ltd., Akanuma 2520, Hatoyama, Saitama 350-0395, Japan}

\author{A.~Tonomura}
\affiliation{Initial Research Project, Okinawa Institute of Science and Technology, c/o Hitachi Advanced Research Laboratory, Akanuma 2520, Hatoyama, Saitama 350-0395, Japan}
\affiliation{Advanced Research Laboratory, Hitachi Ltd., Akanuma 2520, Hatoyama, Saitama 350-0395, Japan}

\author{P.~D.~Brown}
\affiliation{School of Mechanical, Materials and Manufacturing Engineering, University of Nottingham, University Park, Nottingham NG7 2RD, United Kingdom}

\author{R.~P.~Campion}
\affiliation{School of Physics and Astronomy, University of Nottingham, Nottingham NG7 2RD, United Kingdom}

\author{K.~W.~Edmonds}
\affiliation{School of Physics and Astronomy, University of Nottingham, Nottingham NG7 2RD, United Kingdom}

\author{B.~L.~Gallagher}
\affiliation{School of Physics and Astronomy, University of Nottingham, Nottingham NG7 2RD, United Kingdom}

\author{J.~Zemen}

\affiliation{Institute of Physics ASCR, Cukrovarnick\'a 10, 162 53 Praha 6, Czech Republic}

\author{T.~Jungwirth}

\affiliation{Institute of Physics ASCR, Cukrovarnick\'a 10, 162 53 Praha 6, Czech Republic}

\affiliation{School of Physics and Astronomy, University of Nottingham, Nottingham NG7 2RD, United Kingdom}

\date{\today}

\begin{abstract}
We report experimental and theoretical studies of magnetic domain walls in an in-plane magnetized (Ga,Mn)As dilute moment ferromagnetic semiconductor. Our high-resolution electron holography technique provides direct images of domain wall magnetization profiles. The experiments are interpreted  based on microscopic calculations of the micromagnetic parameters and Landau-Lifshitz-Gilbert  simulations. We find that the competition of uniaxial and biaxial magnetocrystalline anisotropies in the film is directly reflected in orientation dependent wall widths, ranging from approximately 40~nm to 120~nm. The domain walls are of the N\'eel type and evolve from  near-$90^{\circ}$ walls at low-temperatures to large angle [1$\bar{1}$0]-oriented walls and small angle [110]-oriented walls at higher temperatures.
\end{abstract}


\maketitle
Magnetic domain walls (DWs) are extensively explored for their potential in integrated memory and logic  devices \cite{Allwood:2005_a,Thomas:2006_a} and because of the number of open basic physics questions related to their dynamics in external magnetic and electric fields \cite{Tatara:2004_a,Zhang:2004_c,Barnes:2005_a,Duine:2007_b,Stiles:2007_a}. Dilute moment ferromagnetic semiconductors, of which (Ga,Mn)As is archetypical, are playing an increasingly important role in this research area \cite{Yamanouchi:2004_a,Wunderlich:2007_c,Duine:2006_a,Nguyen:2006_a}. $p$-type (Ga,Mn)As has a saturation magnetization ($M_s$) which is two orders of magnitude lower than in conventional metal ferromagnets,  while the magnetocrystalline anisotropy energies ($K$) and spin stiffness ($A$) are comparable to the metals \cite{Abolfath:2001_a,Dietl:2001_b,Konig:2001_a,Dietl:2001_c}. The low $M_s$ is due to the dilute Mn moments while the holes in the spin-orbit-coupled valence bands, mediating the long-range  Mn-Mn coupling, produce the large $K$ and $A$.

Among the immediate implications of these characteristics are weak dipolar stray fields which would allow for dense integration of (Ga,Mn)As micro-elements without unintentional cross-links, macroscopic-size domains, square hysteresis loops, and mean-field-like temperature dependent magnetization  \cite{Welp:2003_a,Wang:2005_e}.  The strong in-plane biaxial and uniaxial magnetocrystalline anisotropies \cite{Hrabovsky:2002_a,Welp:2003_a,Wang:2005_e} and weak dipole fields lead also to the formation of unique segmented domain structures and  spontaneous domain  reorganization with changing temperature \cite{Sugawara:2007_a}. An outstanding feature of DW dynamics in (Ga,Mn)As is the  orders of magnitude lower critical current for DW switching than observed for conventional ferromagnets \cite{Yamanouchi:2004_a,Wunderlich:2007_c}.

The basic tool for studying DWs is magnetic structure imaging but here the low saturation moment in (Ga,Mn)As is a problem, greatly reducing the sensitivity of conventional magneto-optical and scanning Hall probe microscopy techniques. For out-of-plane magnetized (Ga,Mn)As films, grown under tensile strain on (In,Ga)As, the resolution achieved with these techniques is limited to $\sim 1\mu$m \cite{Shono:2000_a,Yamanouchi:2004_a,Wang:2007_c} and the sensitivity is further reduced for (Ga,Mn)As layers grown on GaAs under compressive strain with in-plane magnetization \cite{Welp:2003_a,Pross:2004_b,Pross:2004_a}. Imaging internal DW configurations, which are particularly intriguing in the in-plane materials, requires $\sim 100-10$~nm resolution and has therefore remained far beyond the reach of the conventional magnetic microscopy techniques.

In this paper we present the detailed study of in-plane DWs in (Ga,Mn)As. We obtain direct images of DW magnetization profiles and determine the type and width of the DWs and their dependence on the wall orientation and temperature. The high sensitivity to in-plane magnetization is achieved by employing transmission electron microscopy techniques \cite{Hopster:2004} based on the Lorentz deflection of transmitted electrons by the in-plane component of the magnetic induction and on  holographic electron phase retrieval. The interpretation of experiments is based on kinetic-exchange-theory \cite{Jungwirth:2006_a} calculations of  micromagnetic parameters and Landau-Lifshitz-Gilbert (LLG) simulations.

A Ga$_{0.96}$Mn$_{0.04}$As (500~nm)/GaAs (1~nm)/AlAs (50~nm)/buffer-GaAs (100~nm) multilayer was deposited on a GaAs(001) substrate using molecular beam epitaxy. Electron transparent uniform (Ga,Mn)As foils with a wide field of view around 100~$\mu$m were produced by selectively etching away the substrate using the AlAs as a stop layer. The cubic anisotropy favours magnetization along the in-plane $\langle 100 \rangle$ crystalline axes, and the uniaxial anisotropy favours magnetization along one of the $\langle 110 \rangle$ axes. We label the in-plane uniaxial easy axis as the [1$\bar{1}$0] direction, as this was found to be the easy axis in layers grown under similar conditions in the same system \cite{Sawicki:2004_a}. SQUID magnetization measurements on an unetched part of the wafer yield the cubic anisotropy constant $K_{c}$=1.18(0.32)~kJ/$m^{3}$ and the uniaxial anisotropy constant $K_{u}$=0.18(0.11)~kJ/$m^{3}$ at $T$=10(30)~K, and  the Curie temperature $T_c=60$~K.

Electron holography measurements were performed using a 300~kV transmission electron microscope equipped with a cold field-emission electron gun (Hitachi High-Technologies, HF-3300X). The specimen was located at a special stage in between the condenser and the weakly-excited objective lens. In this configuration the internal magnetic structures are expected to be undisturbed by the residual magnetic field, which is estimated to be smaller than 10$^{-4}$~T and oriented perpendicular to the specimen. Electron holography experiments were performed within a near-edge sample region, because of the need for a reference electron wave traveling through vacuum. The electrostatic phase gradient associated with slight thickness variations arising from inhomogeneous chemical etching of the sample near edges was compensated for by subtracting a phase image of the sample in the high temperature paramagnetic state together with a linear wedge correction. The sampling resolution of the CCD camera used for image acquisition was 1.6~nm, whilst the numerical phase reconstruction was performed using a low-pass Fourier mask corresponding to the wavelength of 20~nm for holograms acquired with 5~nm-spacing interference fringes.

\begin{figure}[h]

\hspace*{0cm}\includegraphics[width=.80\columnwidth,bb=0 0 563 473, angle=-0]{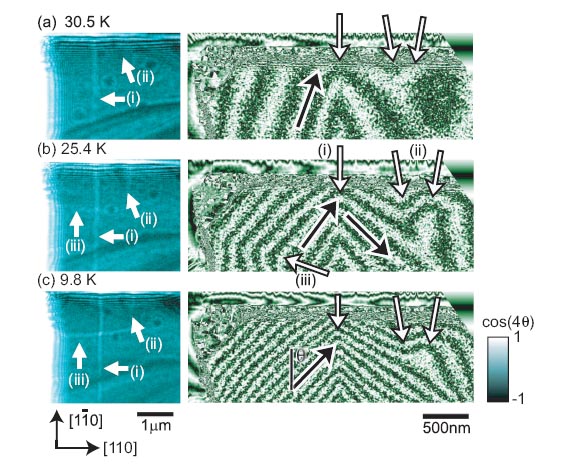}

\vspace*{0cm}
\caption{Lorentz micrographs (left column) and phase images amplified by a factor of four (right column) acquired at (a) 30.5~K, (b) 25.4, and (c) 9.8~K, respectively. Three different types of DWs are observed, marked by white arrows: (i) a wall parallel to [1$\bar{1}$0] that is near-180$^{\circ}$ type at high temperature and near-90$^{\circ}$ type at low temperature; (ii) a pair of DW that is near-180$^{\circ}$ head-on type at high temperature and near-90$^{\circ}$ type at low temperature, and (iii) a near-90$^{\circ}$ type wall parallel to [110] that appeared at low temperature. The local B directions are denoted by black arrows.}
\label{f1}
\end{figure}

Fig.~1 shows the DW phase images (right panels) acquired from a $3\mu{\rm m}\times1\mu{\rm m}$ corner region of the sample foil at 30.5~K,  25.4~K, and 9.8~K, respectively, together with a larger area overviews (left panels) obtained by Fresnel mode Lorentz imaging \cite{Sugawara:2007_a}. The phase ($\phi$) is amplified by a factor of 4 for clarity, and its cosine (i.e. $cos(4 \phi)$) is dislpayed in gray-scale. The magnitude and direction of the magnetic induction ${\bf B}$ are determined separately from the relationship between the phase gradient and magnetic induction,
$\partial\phi/\partial{\rm\bf r}=2\pi e t/h\,{\rm\bf B}({\rm\bf r})$,
where $e$ is the electron charge, $t$ is the film thickness, and $h$ is the Planck constant. The local ${\rm\bf B}$ directions are parallel to the tangent of the equiphase lines, as indicated by black arrows in Fig.~1, and the magnitude of ${\rm\bf B}$ is inverse proportional to the spacing of the equiphase lines.

The Fresnel-mode Lorentz and holographic phase reconstructed images show consistently the location of several DWs by bright contrast lines in the former case and by sharply bent equiphase lines in the latter case. The high resolution electron holography then provides the detailed information on the internal structure of the DWs. As expected   the magnitude of ${\rm\bf B}$ decreases with increasing temperature. The direction of ${\rm\bf B}$ rotates gradually across the wall boundary for all detected DWs implying the  N\'eel type  walls in the studied (Ga,Mn)As film. For the DW denoted as (i) in Fig.~1, ${\rm\bf B}$ rotates from the near [100]/[010] directions towards the [1$\bar{1}$0] direction with increasing temperature, {\em i.e.}, we acquired direct images of a transition from the  near-$90^{\circ}$ in-plane DW at low-temperatures to a near-$180^{\circ}$ wall at higher temperatures. As discussed below this is a demonstration in  DW physics of the competition between cubic and uniaxial anisotropies in the (Ga,Mn)As ferromagnet.

\begin{figure}[h]

\hspace*{0cm}\includegraphics[width=.6\columnwidth, bb=0 0 421 356]{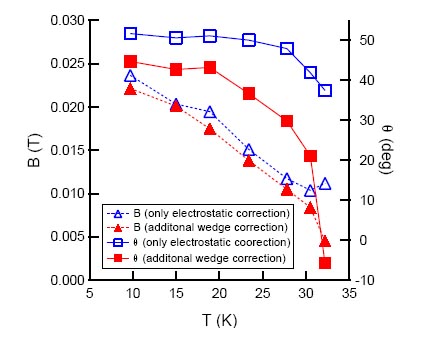}

\vspace*{0cm}
\caption{Magnitude ($|B|$) and direction ($\theta$ ) of $B$ as a function of temperature ($T$).
Only results with subtraction of paramagnetic phase images for thickness variation correction and further linear-wedge correction are displayed.}
\label{f2}
\end{figure}

Fig.~2 summarizes the variation of the magnitude of {\bf B} and the angle $\theta$ of {\bf B} measured from the [1$\bar{1}$0]-axis as a function of temperature for the left end of the DW (i). Filled symbols, corresponding to the phase images in Fig.~1, include linear wedge correction for thickness variation near the sample foil edge along the [1$\bar{1}$0]-direction. For comparison we also plot the uncorrected data which show similar behavior only the variation of $\theta$ appears smaller. We also note that the Lorentz wall contrast observed in the near-edge regions disappeared at temperatures typically 10~K lower than the far-edge region. This suggests that the local temperature near the edges of the sample foil was higher than indicated by the thermal read-out from the liquid helium sample holder, due to restricted or insufficient heat transfer.

\begin{figure}[h]

\hspace*{-0.0cm}\includegraphics[width=0.9\columnwidth, bb=0 0 655 463]{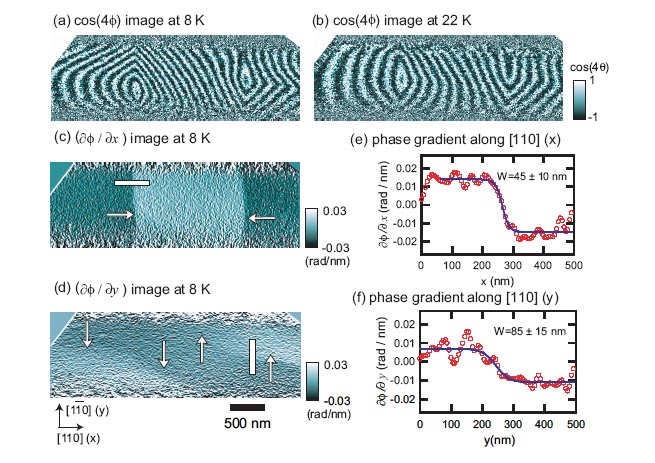}

\vspace*{0cm}
\caption{Phase image amplified by a factor of 4, acquired at 8~K (a) and 22~K (b). $x$-differentiated (c) and  $y$-differentiated (d) images of (a). The DW positions are indicated by white arrows. Projected profile of the phase gradient across the [1$\bar{1}$0] ($x$) -oriented DW (e) and [110] ($y$) -oriented DW (f) along lines indicated by white rectangles.}
\label{f3}
\end{figure}

Phase images  in Fig.~3(a),(b) show vortex-like DWs which clearly demonstrate the dependence of the magnetization rotation angle and width of the DWs on the crystallographic orientation of the wall and temperature. The [110] and [1$\bar{1}$0]-oriented walls evolve from  near-$90^{\circ}$ walls at low-temperatures to a large angle [1$\bar{1}$0]-oriented wall and a small angle [110]-oriented wall at higher temperatures. The width $W_m$ of the walls is obtained by differentiating the phase images with respect to the $x$ ([110]) and $y$ ([1$\bar{1}$0]) directions (see Figs.~3(c)-(f)) and by fitting the measured phase gradient profile with a hyperbolic-tangent function \cite{Hubert:1998_a}. In particular, the $B_y$ profile along the $x$-axis for the narrow [1$\bar1$0]-oriented wall was fitted by
$B_y= B_{y0}\tanh  (2(x-x_0)/W_m)+C$, where
$x_0$, and $C$ are the central position of the wall and a compensation term for the phase gradient, respectively. We obtained  $W_m=45 \pm 10$~nm for the measurement at 8~K and $54 \pm 17$~nm at 22~K. At 30~K the width is, within the error bar, identical to $W_m$ at 22~K. Analogous fitting procedure for the [110]-oriented walls yield $W_m=85 \pm 15$~nm for the measurement at 8~K and $117 \pm 35$~nm at 22~K. At 30~K the [110]-oriented walls have not been resolved.

Quantitatively accurate theoretical modeling of the micromagnetics in (Ga,Mn)As is inherently difficult due to the strong disorder  and experimental uncertainties in  Mn and hole densities.  On a qualitative or semiquantitative level,   the physics underlying the observed  phenomenology of in-plane DWs in (Ga,Mn)As can be consistently described using the well established kinetic-exchange model, as we now discuss in detail. The description is based on the canonical transformation
which for (Ga,Mn)As replaces hybridization of Mn $d$-orbitals with As and Ga $sp$-orbitals by an effective
spin-spin interaction of  $L=0,S=5/2$ local-moments with host valence band states \cite{Jungwirth:2006_a}. The spin-orbit coupling in the band states produces the large magnetocrystalline anisotropies  and, together with the mixed heavy-hole/light-hole character, the large spin stiffness \cite{Abolfath:2001_a,Dietl:2001_b,Konig:2001_a,Dietl:2001_c}.
\begin{figure}[h]

\vspace*{-.0cm}
\hspace*{-.0cm}\includegraphics[width=.95\columnwidth, bb=0 0 653 338]{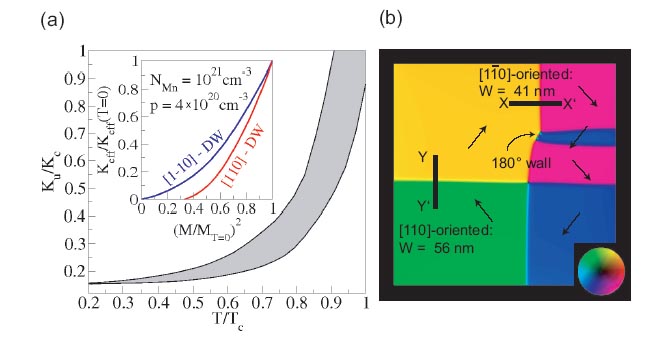}

\vspace*{0cm}
\caption{(a) Microscopic calculations of  $K_u/K_c$ for the whole range of doping parameters considered; inset shows the effective barrier energies for the two walls and the specified hole and local moment densities. (b) LLG simulations for the low temperature micromagnetic parameters of the studied (Ga,Mn)As. Magnetization orientations in the individual domains are highlighted by arrows.}
\label{f4}
\end{figure}

Based on previous detailed characterizations \cite{Jungwirth:2005_b} of the as-grown (Ga,Mn)As materials we assumed in our calculations a range of relevant hole densities, $p=3-4\times 10^{20}$~cm$^{-3}$ and Mn local moment dopings, $x_{Mn}=3-4.5$\% ($N_{Mn}=6-10\times 10^{20}$~cm$^{-3}$). The corresponding mean-field Curie temperatures are between 50 and 100~K, consistent with experiment. First we inspected the theoretical dependence of in-plane magnetocrystalline anisotropies on growth-induced lattice-matching strains. We found that typical strains in as-grown materials have a negligible effect on the in-plane anisotropy energy profiles and, therefore, releasing the strain during the preparation of the thinned, electron-transparent (Ga,Mn)As foil should not affect significantly the properties of in-plane DWs. The in-plane magnetocrystalline anisotropy energy is accurately described by $E(\theta)=-K_c/4 \sin^2 2\theta+K_u \sin^2\theta$, where $\theta$ is measured from the [1$\bar{1}$0] crystal axis. The microscopic origin of the [110]-uniaxial anisotropy component present in most (Ga,Mn)As materials is not known but can be modeled \cite{Sawicki:2004_a} by introducing a shear strain $e_{xy}\sim 0.001-0.01$\%. For the considered range of hole and local moment densities we obtained theoretical $T=0$ values of $K_c$ between approximately 0.5 and 1.5~kJm$^{-3}$, consistent with the low-temperature SQUID measurement of $K_c$. The theoretical $K_c$ values were found to be independent of $e_{xy}$, which is the only free parameter in the theory and whose magnitude and sign was fixed to match the experimental low-temperature $K_u/K_c$ ratio. We then calculated the temperature dependence of $K_u/K_c$ and the corresponding easy-axis angle $\theta=1/2 \arccos(K_u/K_c)$ (the other easy-axis is placed symmetrically with respect to the [1$\bar{1}$0]-direction.) The results, shown in Fig.~4(a), are fairly universal for all $p$'s and $N_{Mn}$'s considered in the calculations and consistent with experimental data in Fig.~2.

An order-of-magnitude estimate of the theoretical DW width is given \cite{Konig:2001_a} by the length-scale $\sqrt{A/K_{eff}}$, where $K_{eff}$ is the effective anisotropy energy barrier separating the bistable states on respective sides of the DW. For $N_{Mn}\approx 10^{21}$~cm$^{-3}$, the mean low-temperature value of $K_{eff}\approx K_c/4\approx 0.3$~kJm$^{-3}$ (with 20\% variation in the considered range of hole densities) and the spin-stiffness  $A\approx 0.4$~pJm$^{-1}$ (nearly independent of $p$), yielding typical DW width $W_m\sim 40$~nm, in agreement with experiment. Despite the relatively small $K_u/K_c\approx 0.15$ and the corresponding small tilt by $\approx 10^{\circ}$ of the easy-axis from the [100]/[010] directions at low temperatures, $K^{[1\bar{1}0]}_{eff}\approx K_c/4+K_u/2$ for the larger-angle, [1$\bar{1}$0]-oriented DW is already about twice as large as $K^{[110]}_{eff}\approx K_c/4-K_u/2$ for the smaller-angle, [110]-oriented DW (see Fig.~4(a)), explaining the sizable difference between the respective experimental DW widths at 8~K. The observed temperature dependence of $W_m$ can be qualitatively understood by considering the approximate magnetization scaling of $K_c\sim M^4$, $K_u\sim M^2$, and $A\sim M^2$ \cite{Wang:2005_e,Dietl:2001_c}. This implies for the [1$\bar{1}$0]-oriented DW that $W_m$ initially increase with temperature and then saturates at high $T$ while the [110]-wall width steadily increases with $T$, becoming unresolvable at $K_u(T)/K_c(T)\approx 1/2$. (Note that a more quantitative discussion of the temperature dependence of $W_m$ is also hindered by the relatively large experimental error bars for this quantity.) 

Magnetic dipolar fields play a marginal role in the dilute moment (Ga,Mn)As ferromagnets which may explain, together with the $<180^{\circ}$ wall-angle, the N\'eel type of the observed DWs despite their relatively small thickness. We further elaborate on this and on the above qualitative arguments by performing the  LLG simulations using micromagnetic parameters of the studied (Ga,Mn)As material. These calculations, shown in Fig.4(b) for $T=8$~K, confirm the N\'eel type of the DWs, the evolution from  near-$90^{\circ}$ walls at low-temperatures to larger angle [1$\bar{1}$0]-oriented walls and smaller angle [110]-oriented walls at higher temperatures, and the increasing anisotropy of the DW widths with increasing temperature with $W_m$ ranging from 40 to 100~nm. This leads us to the conclusion that our combined experimental and theoretical work represents an important extension to our understanding of the micromagnetics of in-plane magnetized (Ga,Mn)As down to the smallest relevant length-scale, the individual DW width. Since our findings are likely unaffected by the constraints of the experimental technique on the lateral and vertical sample dimensions, our approach has a generic utility as a basis for DW studies in dilute moment ferromagnets.

We acknowledge valuable discussions with M.R. Scheinfein and T. Yoshida and support from  EU Grant  IST-015728, from UK Grant GR/S81407/01, from CR  Grants 202/05/0575, 202/04/1519, FON/06/E002, AV0Z1010052, and LC510.


\end{document}